\documentclass{article}

\usepackage{arxiv}

\usepackage[utf8]{inputenc} % allow utf-8 input
\usepackage[T1]{fontenc}    % use 8-bit T1 fonts
\usepackage{hyperref}       % hyperlinks
\usepackage{url}            % simple URL typesetting
\usepackage{booktabs}       % professional-quality tables
\usepackage{amsfonts}       % blackboard math symbols
\usepackage{nicefrac}       % compact symbols for 1/2, etc.
\usepackage{microtype}      % microtypography
\usepackage{lipsum}		% Can be removed after putting your text content
\usepackage{graphicx}
\usepackage{natbib}
\usepackage{doi}

\usepackage{amsmath}
\usepackage{amsfonts}
\usepackage{amssymb}
\usepackage{amsbsy}
\usepackage{amsthm}
\usepackage[nolist]{acronym}
\usepackage{eurosym}
\usepackage[table]{xcolor}
\usepackage{cleveref}
\usepackage{subcaption}

\title{Economy and sustainability analysis with a novel modular configurable multi-modal white-box building model}

\date{} 					% Or removing it

\author{ \href{https://orcid.org/0009-0005-0067-8019}{\includegraphics[scale=0.06]{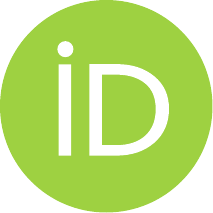}\hspace{1mm}Haozhen~Cheng} \\
	Institute for Automation and Applied Informatics\\
	Karlsruhe Institute of Technology\\
	76131 Karlsruhe, Germany \\
	\texttt{haozhen.cheng@kit.edu} \\
        \And
	\href{https://orcid.org/0000-0002-3572-9083}{\includegraphics[scale=0.06]{orcid.pdf}\hspace{1mm}Veit~Hagenmeyer} \\
	Institute for Automation and Applied Informatics\\
	Karlsruhe Institute of Technology\\
	76131 Karlsruhe, Germany \\
	\texttt{veit.hagenmeyer@kit.edu} \\
        \And
        \href{https://orcid.org/0000-0002-1463-7606}{\includegraphics[scale=0.06]{orcid.pdf}\hspace{1mm}Hüseyin K.~Çakmak} \\
	Institute for Automation and Applied Informatics\\
	Karlsruhe Institute of Technology\\
	76131 Karlsruhe, Germany \\
	\texttt{hueseyin.cakmak@kit.edu} \\
}

\renewcommand{\shorttitle}{Economy and sustainability analysis with a novel building model}

\hypersetup{
pdftitle={Economy and sustainability analysis with a novel modular configurable multi-modal white-box building model},
pdfsubject={eess.SY},
pdfauthor={Haozhen~Cheng, Veit~Hagenmeyer, Hüseyin K.~Çakmak},
pdfkeywords={Building, Modelica, Co-simulation, Sustainability, Economy},
}

\begin{document}

\begin{acronym}
    \acro{KIT}{Karlsruhe Institute of Technology}
    \acro{FZJ}{Forschungszentrum Jülich}
    \acro{DH}{District Heating}
    \acro{TABULA}{Typology Approach for Building Stock Energy Assessment}
    \acro{DHW}{Domestic Hot Water}
    \acro{HIU}{Heat Interface Unit}
    \acro{HVAC}{Heating, Ventilation and Air Conditioning}
    \acro{FMI}{Functional Mock-up Interface}
    \acro{FMU}{Functional Mock-up Unit}
    \acro{LPG}{LoadProfileGenerator}
    \acro{PV}{Photovoltaic}
\end{acronym}

\maketitle

\begin{abstract}
    This paper presents a novel modeling approach for building performance simulation, characterized as a white-box model with a high degree of modularity and flexibility, enabling direct integration into complex large-scale energy system co-simulations. The introduced model is described in detail, with a focus on its modular structure, and proposes various configurations that include various building insulation, heating methods, occupancy patterns, and weather data to analyze different scenarios, and the energy consumption, \mbox{$CO_2$} emissions, and heating costs are compared and analyzed across 36 introduced scenarios. The thermodynamic behavior of the model is shown to be consistent with real-world conditions, and the comparison of the scenarios concludes that the use of heat pumps for indoor heating in well-insulated buildings has significant economic and sustainability benefits, whereas the use of natural gas-fueled boilers is more cost-effective for buildings with low energy ratings.
\end{abstract}

% keywords can be removed

\section{Introduction}
\label{sec:intro}

In today's energy system field, the digitalization process provides key assistance in achieving the sustainable development of energy systems \citep{Singh2022}. Among them, digital twins of energy systems have multiple advantages; in particular, they can help in decision making more effectively by evaluating and predicting various scenarios based on existing data. In addition, a qualified digital twin model can provide optimization strategies for existing energy systems. In the face of increasing energy demands and society striving to reduce the impacts of climate change \citep{Klimaschutzgesetz2024}, the search for more sustainable and efficient energy systems has become a priority. To design district energy digital twins, it is necessary to develop effective and easy-to-use building and energy network models. This is because buildings, as the basic units of the energy system, are not only energy consumption terminals, their thermodynamic and economic behaviors also largely determine the living comfort of residents and their economic expenditure on energy.

As for previous works on building modeling, Johari et al. \citep{johari2022} evaluated several building configurations using different building energy models \citep{ciccozzi2023} with modeling tools such as IDA ICE (\textit{equa.se/de/ida-ice}), TRNSYS (\textit{trnsys.com}), and EnergyPlus (\textit{energyplus.net}), and analyzed the applicability of these tools for energy analysis of different types of buildings. Based on three types of buildings presented from the \ac{TABULA} \citep{tabula2016} project, i.e. single house, terrace house and multifamily house, Bruno et al. \citep{bruno2016} analyzed the simplified thermodynamic behavior of the buildings using the TRNSYS tool, and focused on analyzing simplified building envelope models and individual building heat loads.

In this paper, the authors propose a new approach for building modular modeling using the Modelica language (\textit{modelica.org}). Although Modelica is not a mainstream tool for modeling building energy demand \citep{alfalouji2023}, it provides modular and visual modeling possibilities using the Dymola \citep{Dymola3ds2024} modeling tool. This language is open source with a large and rapidly growing community in different industrial application areas and academia \citep{Schweiger2017}, and its non-causal characteristics make it simpler and clearer to model complex systems than other equation-based modeling languages (such as Simulink). Qiu et al. \citep{Qiu2024} mentioned that there is no need to delve into complex algebraic equations when modeling high-level components, which shows great potential for modeling large-scale complex systems. In addition, Perera et al. \citep{perera2016} developed a multilayer building heating model in MATLAB (\textit{mathworks.com}) and Modelica environments, and showed that Modelica models are more robust.

The contribution of this paper is a new multi-physics/modal and flexibly configurable modular building model using the Modelica language, which is evaluated with an in-depth benchmarking of economics and sustainability in the context of house heating. The introduced model allows for the utilization of the setup of complex district energy digital twins. As a white-box model, it has sufficient degrees of freedom to observe and control all parameters of the devices inside the building. Based on its modular and easy-to-configure characteristics, the authors merge and divide the modules in the model into three large categories in this study, and multiple variants are created for each category to simulate and analyze the energy consumption and \mbox{$CO_2$} emissions of buildings with different configurations on thermal insulation and heating methods under different climates and occupancy behaviors. Thus, this lays the foundation for the analysis of district energy systems in special or extreme situations in the future, such as natural gas shortages, heat waves, or other destructive events.

The remainder of this paper is structured as follows:~\Cref{sec:modular building model} introduces the modular building model with a detailed description of each part, whereas~\Cref{sec:case study} presents the analyzed cases. The simulation results are compared and analyzed for each case in~\Cref{sec:evaluation}, and a brief summary and outlook are provided in~\Cref{sec:conclusion}.

\section{Modular Building Model}
\label{sec:modular building model}
\begin{figure}[ht]
     \centering
     \subcaptionbox{The overall appearance of the building model in Dymola.\label{subfig:building model entire}}{
         \includegraphics[width=0.45\textwidth]{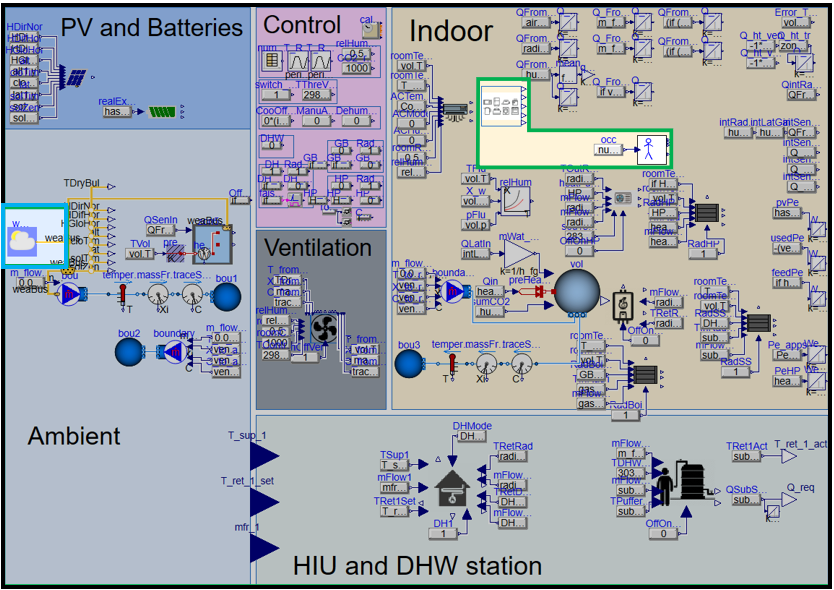}
     }
     \hspace{8pt}
     \subcaptionbox{The three parts of the model from~\Cref{subfig:building model entire} (Part-1: black, Part-2: green, Part-3: blue).\label{subfig:3 parts}}{
         \includegraphics[width=0.35\textwidth]{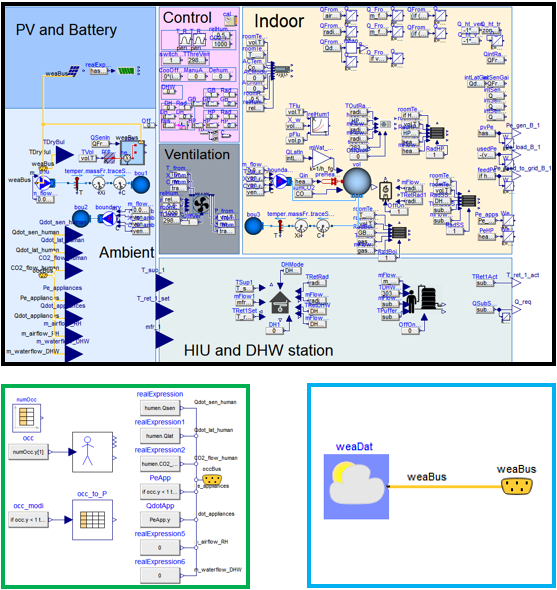}
     }
     \caption{The modular building model in Dymola and its three parts.}
     \label{fig:building model}
\end{figure}

As mentioned in~\Cref{sec:intro}, the model proposed by the authors in the present paper is implemented in the Dymola simulation software using the Modelica modeling language. Its overall appearance in Dymola is shown in~\Cref{subfig:building model entire}. 
From the annotations, it can be seen that the building model has a clear modular feature, including the external environment (Ambient) module, Ventilation module, Indoor module, \ac{PV} and Battery module, Control module, and the Heat grid interface module that can be combined with the heat grid model. Each module may contain multiple submodules. For example, the indoor module contains heat pump, gas boiler, radiator and other equipment modules, whereas the heat grid interface module contains \ac{HIU} and \ac{DHW} modules.

In the present study, the building model shown in~\Cref{subfig:building model entire} is divided into three parts:
\begin{itemize}
\item{Part-1: Modules, which represent building envelope and indoor equipment,}
\item{Part-2: Modules, which represent resident occupancy behavior, and}
\item{Part-3: Modules, which represent weather,}
\end{itemize}
which are shown in~\Cref{subfig:3 parts}, and the position of each part is marked in~\Cref{subfig:building model entire} using various colors. The submodules contained in each part, and how these three parts are connected, are introduced in detail in the remainder of this section.

\subsection{Part-1: Building Envelope and Indoor Equipment}
As can be seen from~\Cref{fig:building model}, most of the modules of the model are divided into part-1, which not only includes the building envelope model and ventilation model, which have a key impact on the thermodynamic behavior of the building, but also includes the heating device model, which is crucial to the comfort of residents. In this section, several modules related to the case analysis in the present paper are described in detail.

\subsubsection{Building Envelope}
\begin{figure}[htb]
{
\centering
\includegraphics[width=0.7\textwidth]{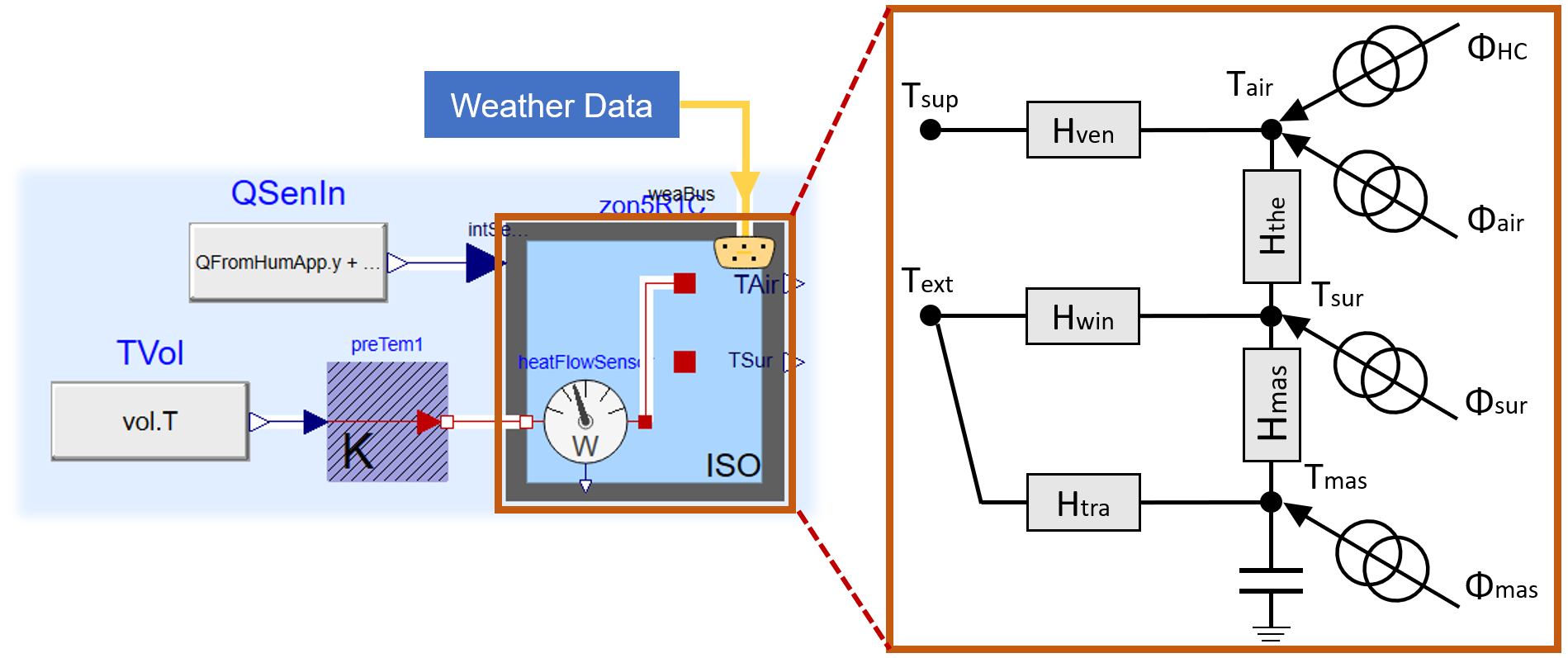}
\caption{The building envelope model and its electrical 5R1C schematic diagram \citep{bruno2016}.}\label{fig:envelope}
}
\end{figure}

It is important to use a reasonable building envelope model to calculate a building’s heating or cooling demand. In the Modelica \textit{Buildings} library \citep{WetterZuoNouiduiPang2014}, the developers of the library established a simple, clear, and easy-to-understand mathematical model, namely the \textit{5R1C} building envelope model (see the right-hand side of~\Cref{fig:envelope}), which is described in the EN ISO 13790:2008 standard \citep{eniso2008}. This model is analogous to the resistance and capacitance in the circuit model and simultaneously considers the characteristics of heat conduction and heat capacity in buildings through five \textit{conductance} elements representing the heat conduction characteristics, five \textit{voltage nodes} representing the temperature, and one \textit{capacitance} element representing the building heat capacity of the entire zone.

The three temperature nodes used to characterize the temperature inside the building are the indoor air temperature \mbox{$T_{air}$}, internal surface temperature of the envelope \mbox{$T_{sur}$}, and area mass temperature \mbox{$T_{mas}$}, which are thermally connected to two temperature nodes located outdoors, with \mbox{$T_{sup}$} representing the supply air temperature and \mbox{$T_{ext}$} representing the outdoor air temperature through various thermal conductance elements. Among them, the window heat transfer \mbox{$H_{win}$} and the heat transfer of the opaque component \mbox{$H_{op}$} (divided into \mbox{$H_{tra}$} and \mbox{$H_{mas}$}) determine the indoor thermal transmission characteristics with outside. \mbox{$H_{mas}$} is the thermal conductance between the surface and the mass nodes and is defined as:
\begin{equation}
\label{eq: Hmas}
H_{mas} = h_{ms} \cdot f_{ms} \cdot A_{f}
\end{equation}
where \mbox{$h_{ms}$} is the heat transfer coefficient between the two nodes, with a fixed value of  9.1$W/(m^{2}K)$, $f_{ms}$ is a correction factor, which can be assumed as 2.5 for light and medium building constructions, and 3 for heavy, and $A_{f}$ is the floor area. Because the transmission coefficients for opaque and glazing elements $H_{op}$ and $H_{win}$ are defined by their respective U-values and areas, the expression for $H_{tra}$ can be derived as follows:
\vskip -8pt
\begin{equation}
H_{tra} = \frac{1}{\frac{1}{H_{op}} - \frac{1}{H_{mas}}}
\end{equation}

In addition, the heat transfer characteristics between the indoor air and inner surface are expressed by $H_{the} = h_{as} \cdot A_{tot}$, where $h_{as}$ is the heat transfer coefficient between the air node and surface node, with a fixed value of 3.45$W/(m^{2}K)$, and $A_{tot}$ is the entire area of all surfaces facing the air-conditioned building area, which is often chosen as a fixed multiple ($rat_{sur}$) of the floor area $A_{f}$.

As a coupling element between the interior and exterior environments of a building, the building envelope model uses weather data, indoor air temperature, and internal heat gain as inputs to simulate the thermodynamic characteristics of the building together with other modules.

\subsubsection{Heat Pump}
As a key component in energy systems, heat pumps have great potential to save energy and improve overall energy efficiency \citep{Chua2010}. Through the coordination of a compressor, an expansion valve, heat exchangers, refrigerant, and control elements, it is possible to absorb heat from or release it to the external environment, thus achieving a higher efficiency than with thermal resistance heating equipment, making a significant contribution to reducing $CO_{2}$ emissions \citep{Bagarella2016}.

\begin{figure}[htb]
{
\centering
\includegraphics[width=0.9\textwidth]{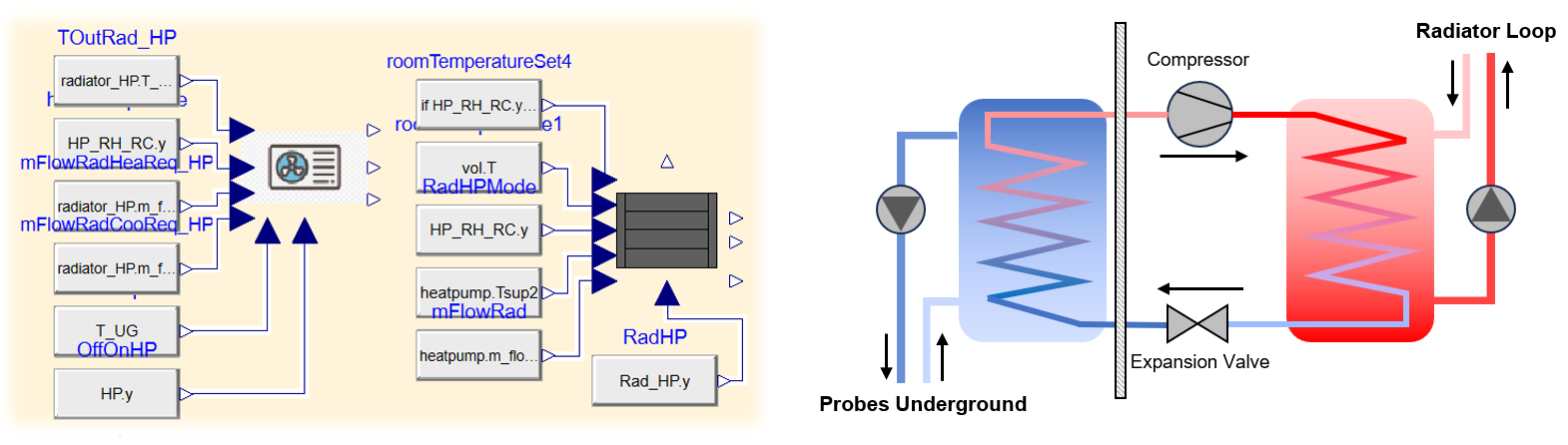}
\caption{The appearance of the heat pump and radiator heating in the Modelica building model (left) and the schematic diagram of the ground source heat pump principle (right).}\label{fig:heatpump}
}
\end{figure}

Heat pumps have been used in many applications such as air conditioners and refrigerators. Most indoor air conditioners are air-source heat pumps. In winter, the temperature of the refrigerant flowing through the expansion valve is lower than the outside temperature, so that it can absorb heat from the outside air. After flowing through the compressor, the absorbed heat is transferred to the room through the heat exchanger to achieve indoor heating. For ground-source heat pumps, which are generally used for indoor radiators or floor heating, the low-temperature refrigerant after flowing through the expansion valve tends to absorb heat from the underground because the underground temperature is normally higher than the surface and is relatively stable in winter. The model appearance and principle diagram are presented in~\Cref{fig:heatpump}.

The heat pump uses the inlet water temperature and the water flow rate on the load and source sides, and some control signals (such as cooling or heating control signals) as input signals, and the outlet water temperature and power consumption as outputs, coupled with the radiator model (the right module in the left figure of~\Cref{fig:heatpump}) to provide the required thermal power for the indoor environment. In this paper, only the use of ground-source heat pumps for heating in winter is discussed, and air-source heat pumps (air conditioners) and summer cooling are not considered for the time being.

\subsubsection{Gas Boiler}
Over the past few decades, gas-fired heating boilers have been one of the most common sources of residential heating demand \citep{Simic2021}, and 36\% of the space heating demand is met by using natural gas as the fuel for gas boilers \citep{Eurostat2019}.

\begin{figure}[htb]
{
\centering
\includegraphics[width=0.9\textwidth]{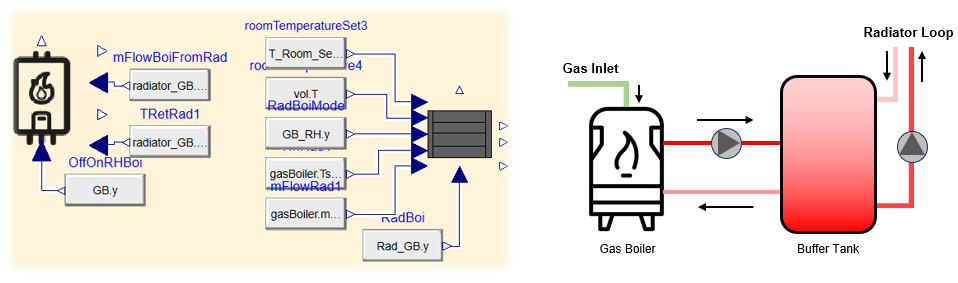}
\caption{The appearance of the gas boiler and radiator heating in the Modelica building model (left) and the schematic diagram of them (right).}\label{fig:gasboiler}
}
\end{figure}

Similar to a heat pump, the water returned from the radiator is heated by the gas boiler and then returned to the radiator to distribute heat indoors. In the model in the present paper, the authors add a buffer tank between the gas boiler and the radiator. The purpose of this buffer tank is to buffer the heat supply from the boiler, thus optimizing operation when demand on the user side changes rapidly to avoid the boiler switching on and off too frequently. At the same time, it ensures that a reserve of heat energy is always available, even at peak loads in the heat demand. It is worth noting that, unlike the heat pump model, which outputs electricity consumption, one of the gas boiler's output is natural gas consumption. The model outline and working diagram are presented in~\Cref{fig:gasboiler}.

\subsubsection{PV and Batteries}

As the demand for social energy continues to grow, solar energy has attracted widespread attention as an effective alternative to conventional energy owing to its renewable, low-cost, and safe properties \citep{Fazal2023}. In particular, when a \ac{PV} system is combined with an energy storage system, it can effectively balance the volatility and unpredictability of \ac{PV} power generation \citep{Li2023}.

\begin{figure}[htb]
{
\centering
\includegraphics[width=0.9\textwidth]{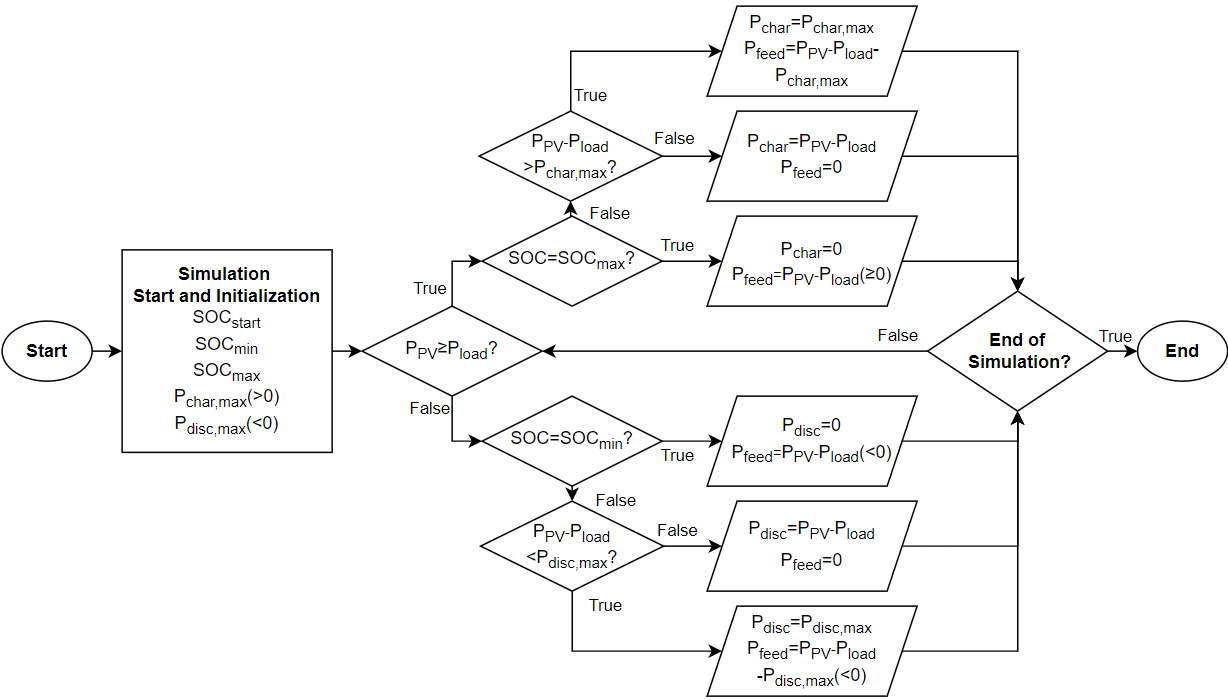}
\caption{Signal flow graph of the relationship between battery, \ac{PV}, household load and power grid. $P_{load}$ means the power consumption of load, $P_{PV}$ is the \ac{PV} power generation, while $P_{feed}$ indicates the power fed into the power grid (positive means fed in, negative means drawn).}\label{fig:battery}
}
\end{figure}

The present study assumes that the electricity generated by the \ac{PV} system can be used not only for household appliances and heat pumps, but that the surplus electricity can also be stored in batteries or profitably fed into the grid. The control flow graph of the power storage and feeding is shown in~\Cref{fig:battery}. The model also considers the maximum charging and discharging power of the battery $P_{char,max}$ and $P_{disc,max}$ to ensure an adequate service life.

\subsection{Part-2: Occupancy}
\label{subsec:occupancy}
The energy system serves human beings, and human behavior in turn greatly affects the operation of the energy system. Occupants are responsible for the energy performance of a building \citep{Gaetani2016}. With the optimization of building envelopes, technological advances, and the popularization of low-energy systems, occupant behavior has an ever-increasing influence on building performance \citep{Clevenger2006}. However, owing to the diversity of households, the composition of the residents, and the regions, it is very difficult to accurately predict occupant behavior.

Occupancy models are usually divided into presence and action models, with action models representing various user behaviors in terms of regulation (e.g., opening and closing windows) and operation (e.g., using household appliances). To make the model more convincing, the authors set the presence and action models in a linear relationship, as shown in Table 1.

\begin{table}[htb]
\caption{Statistics electricity consumption of large families of different sizes in Germany \citep{Verivox2024}.}\label{tab:first}
\centering
\begin{tabular}{ccc}
\hline
Households & Annual electrical consumption [kWh] & Average [W]\\ \hline
1-person-household & 1600 & 182.65\\
2-person-household & 2900 & 331.05\\
3-person-household & 3700 & 422.37 \\
4-person-household & 4250 & 485.16\\
\hline
\end{tabular}
\end{table}
According to the electricity statistics for large families in Germany, the average annual electricity consumption of families with different numbers of people is shown in the second column of the table. In the present study, it is assumed that when performing the time series simulation, the power consumption of the building in each time step is related to the number of people in the building in this step. For example, at time step N, if there are two people in the building, then $P_{t=N} = 331.05W$, which is the value corresponding to the third column in the table. In addition, the standby power of appliances accounts for an average of 10 to 20 percent of the total household power consumption \citep{Kocos2021}, therefore, when no one is home, the power consumption is assumed to be 15 percent of the power consumption when the household is fully occupied. Note that the data in the table are general electricity consumption data excluding heat pumps and air conditioners. The details of the heat pump consumption are discussed further in~\Cref{sec:case study} and~\Cref{sec:evaluation}.

\begin{figure}[ht]
     \centering
     \subcaptionbox{The presence profiles in a large family of 1 to 4 during the week and at the weekend (WD: weekdays, WE: weekend).\label{subfig:ref presence}}{
         \includegraphics[width=0.53\textwidth]{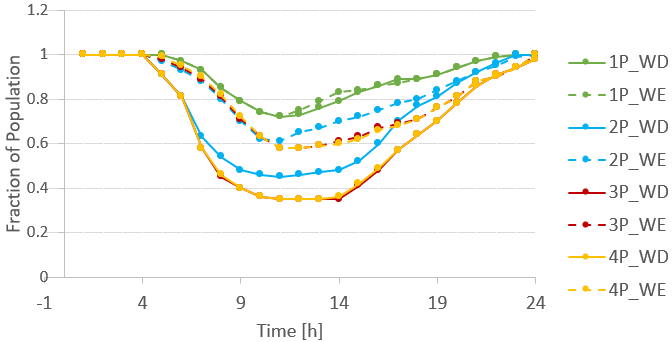}
     }
     \hspace{4pt}
     \subcaptionbox{Comparison of the average of the presence profiles generated using binomial sampling with reference values for 3-person-household (top: weekdays, bottom: weekends).\label{subfig:presence comparison}}{
         \includegraphics[width=0.42\textwidth]{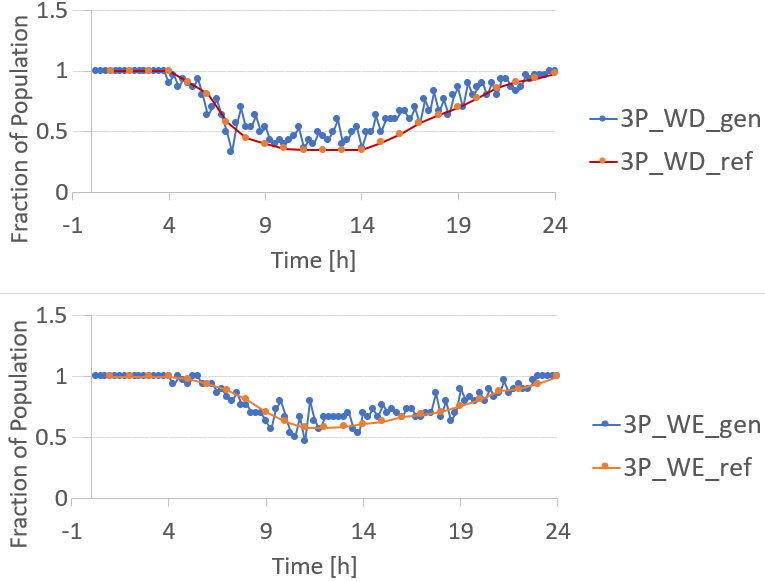}
     }
     \caption{Referenced and generated presence profiles.}
     \label{fig:presence}
\end{figure}

Based on these assumptions, obtaining a reasonable presence model is the core issue of this part. Occupancy patterns vary for different household sizes and even at different times of the week (weekdays and weekends). In general, the profile curves on weekends are smoother than those on weekdays, and occupancy rates are lower during the day than at night. After analyzing and summarizing publicly available datasets of occupancy patterns, Mitra et al. \citep{Mitra2020} created occupancy presence models for four household sizes on weekdays and weekends by combining the average schedules for each (see~\Cref{subfig:ref presence}). The authors then used the \textit{binom} library in Python to generate the presence profiles for each household using binomial random sampling. To verify the validity of this method, the authors sample a three-person family 10 times on weekdays and on weekends, and averaged them. The results of the comparison between the generated presence profiles and the reference curves given in~\Cref{subfig:ref presence} are shown in~\Cref{subfig:presence comparison}, which show that the presence model created in this way generally conforms to the reference model and ensures a certain degree of randomness.

According to the law of conservation of energy, almost all of the electrical consumption of appliances in a household is eventually converted into heat and dissipated into the room environment. Therefore, the authors assume that the thermal power of the electrical load is equal to its power consumption. In addition, to calculate the impact of human activities on the thermodynamic behavior of buildings, the body's own heat production should not be ignored. The heat dissipated by the human body to its surroundings is not a constant value and is affected by many factors, such as ambient temperature, air flow rate, clothing, and individual physiological differences. Generally, the average value of sensible heat power generated by the human body is approximately 120$W$ \citep{ETB2004}. This constant value is used in the present study as the heat production power of the human body for the simulation.

\subsection{Part-3: Weather}
In the weather module, local weather data are converted into \textit{.mos} format (\textit{energyplus.net/weather}) and imported the corresponding weather data into the Modelica models through the \textit{ReaderTMY3} component in the Modelica \textit{Buildings} library. The names of each weather data, their respective descriptions, and the other modules they are used in are shown in~\Cref{tab:weather data}.

\begin{table}[htb]
\caption{Description of the parameters of weather data and where they are used.}\label{tab:weather data}
\centering
\begin{tabular}{ccccc}
\hline
Output Parameter & Description & PV & Envelope & Ventilation\\ \hline
HDirNor & Input direct normal radiation [$W/m^2$] & x & x & \\
HDifHor & Input diffuse horizontal radiation [$W/m^2$] & x & x & \\
HGloHor & Input global horizontal radiation [$W/m^2$] & x & x & \\
alt & Altitude of the location [$m$] & x & x & \\
cloTim & One-based day number [$s$] & x & x & \\
lat & Latitude of the location [$rad$] & x & x & \\
solTim & Solar time [$s$] & x & x & \\
solZen & Zenith angle of sun [$rad$] & x & x & \\
pAtm & Atmospheric pressure [$Pa$] &  &  & x \\
TDryBul & Dry bulb temperature at ground level [$K$] &  & x & x \\
relHum & Relative humidity &  &  & x \\
\hline
\end{tabular}
\end{table}

\section{Case Study}
\label{sec:case study}
Combining the modular and parameterizable features of the proposed model in the present paper, the authors consider each part of the model as a basic model and extend them into multiple variants through parameter tables. A brief description of each variant is provided in~\Cref{tab:variants}. In the simulation analysis phase, each variant of these three parts is combined, forming a total of $6\times3\times2 = 36$ scenarios. Under the same indoor temperature control strategy, the authors summarize and compare the energy consumption and equivalent $CO_2$ of each scenario.
\begin{table}[htb]
\caption{Variants definition of each part.}\label{tab:variants}
\centering
\begin{tabular}{lll}
\hline
\multicolumn{1}{c}{Part-1 (Building configuration)} & \multicolumn{1}{c}{Part-2 (Occupancy)} & \multicolumn{1}{c}{Part-3 (Weather)} \\ \hline
\multicolumn{1}{l}{B1: Worse insulation + Gas boiler} & \multicolumn{1}{c}{O1: Normal} & \multicolumn{1}{c}{W1: Normal} \\
\multicolumn{1}{l}{B2: Worse insulation + HP} & \multicolumn{1}{l}{} & \multicolumn{1}{l}{} \\
\multicolumn{1}{l}{B3: Worse insulation + HP + PV and Battery} & \multicolumn{1}{c}{O2: Vacation} & \multicolumn{1}{l}{} \\
\multicolumn{1}{l}{B4: Better insulation + Gas boiler} & \multicolumn{1}{l}{} & \multicolumn{1}{c}{W2: Cold wave} \\
\multicolumn{1}{l}{B5: Better insulation + HP} & \multicolumn{1}{c}{O3: Corona-time} & \multicolumn{1}{l}{} \\
\multicolumn{1}{l}{B6: Better insulation + HP + PV and Battery} & \multicolumn{1}{l}{} & \multicolumn{1}{l}{} \\
\hline
\end{tabular}
\end{table}

The thermal insulation of a building envelope refers to the ability of the envelope (e.g., walls, roofs, and windows) to prevent the transfer of heat from the inside to the outside or from the outside to the inside, and it is an important factor affecting the thermal comfort and energy efficiency of a building. With improvements in residents' living comfort requirements, buildings with good thermal insulation performance can theoretically reduce heating and cooling energy consumption and improve the energy utilization efficiency of buildings. The key factors affecting the thermal insulation of a building include the U-value, which takes into account the thermal conductivity and thickness of the building materials, window air tightness, thermal bridge effects, etc. In the present study, the authors use building data from the TABULA dataset \citep{tabula2016} with code \textit{DE.N.SFH.05.Gen} to classify the thermal insulation of the building envelope as "Worse" and "Better" that corresponds to the \textit{Existing State} of the building and the state of the building after \textit{Usual Refurbishment} in TABULA respectively.

Each module in the model proposed in the present paper can be turned off or on by setting 0 or 1 in the parameter table, thus the authors propose two ways of heating the hot water in the indoor radiator: heating by gas boiler (using natural gas) and heating by heat pump. Because the operation of a heat pump consumes more electricity, the authors will also show the results of combining a heat pump with a \ac{PV} battery system. The battery capacity is set to 10$kWh$, the maximum charge/discharge rate is 5$kW$, and the area of the photovoltaic panels is set to half of the building footprint, which is 84.45$m^2$.

To ensure comparability of the results from each scenario, the indoor temperature of the building is controlled during the heating season using the following strategy: when the indoor temperature was lower than 20°C, the heating equipment starts to operate until the indoor temperature was higher than 22°C, after which the heating equipment stops operating. In addition, the water flow rate through the radiator is influenced by the actual indoor temperature and the set temperature, which is 20°C at night (after 11:00 p.m. and before 6:00 a.m.) and 22°C during the day (after 7:00 a.m. and before 10:00 p.m.).

In the configuration of building occupancy, the authors assume three different occupancy patterns for a single family house with four family members: a \texttt{\small Normal} occupancy pattern, in which the occupancy curve matches the reference curve in~\Cref{subsec:occupancy}; a \texttt{\small Vacation} pattern, in which the room is always unoccupied; and a \texttt{\small Corona-time} pattern, in which the room is always full.

In addition, the weather data \texttt{\small Normal} used here are from the year 2023, measured in the Living Lab at \ac{KIT} Campus North and stored in a time series database with a Grafana (\textit{grafana.com}) interface \citep{hagenmeyer2016}. The \texttt{\small Cold Wave} weather data used in the present study were modified from the \texttt{\small Normal} weather data by lowering the temperature by 4°C.

\section{Evaluation}
\label{sec:evaluation}
\begin{figure}[htb]
{
\centering
\includegraphics[width=0.6\textwidth]{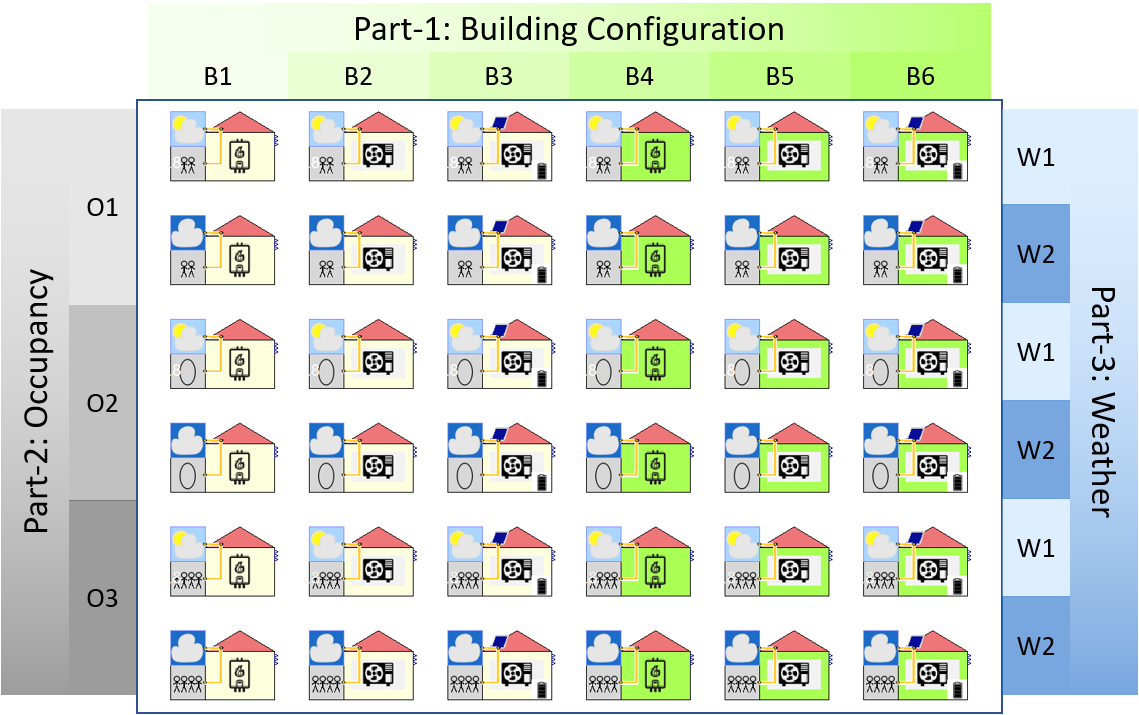}
\caption{36 scenarios enumerated according to the permutations of the three parts.}\label{fig:36 variants}
}
\end{figure}

In this section, the authors first illustrate the correctness and feasibility of the model by observing several variables, such as the indoor temperature, and then provide a detailed enumeration and analysis of the simulations of all the scenarios, and conclude with a discussion of these results. All 36 scenarios to be analyzed and compared in this section are shown in~\Cref{fig:36 variants}. The codes of different combinations are expressed as combinations of the codes of their parts. For example, \texttt{\small B1O1W1} indicates that the building is poorly insulated and equipped with a gas boiler, with normal occupancy behavior and normal weather data as input.
 
\subsection{Indoor Temperature}
\begin{figure}[ht]
     \centering
     \subcaptionbox{Indoor temperatures in building B1O1W1 and B1O1W2, and outdoor temperature W1.\label{subfig:temp_weather}}{
         \includegraphics[width=0.47\textwidth]{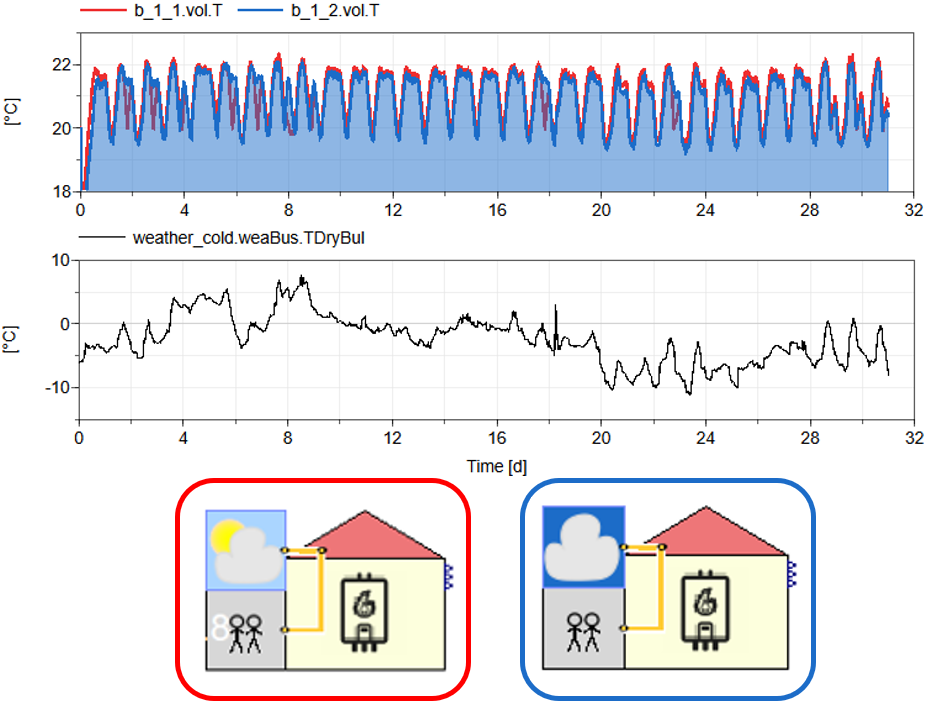}
     }
     \hspace{5pt}
     \subcaptionbox{Indoor temperatures in building B1O1W1 and B2O1W1, and outdoor temperature W1.\label{subfig:temp_GBHP}}{
         \includegraphics[width=0.47\textwidth]{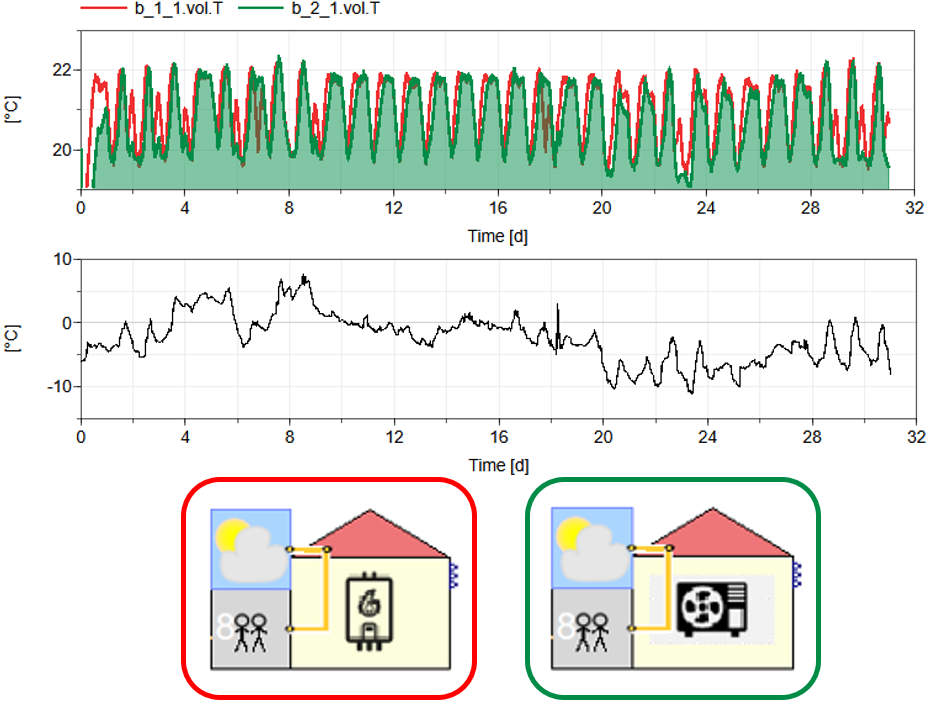}
     }
     \caption{Changes in indoor temperature over a month in buildings B1O1W1 (red), B1O1W2 (blue) and B2O1W1 (green) at different outdoor temperatures and with different heating equipment.}
     \label{fig:temp}
\end{figure}

To observe whether the modeling of the building envelope and heating elements, as well as the fit between the various component modules, meets expectations, the authors first observe the indoor temperatures of the three buildings \texttt{\small B1O1W1}, \texttt{\small B1O1W2}, and \texttt{\small B2O1W1} over a one-month period (January) with a step size of 15 minutes. The results of these three simulations are presented in~\Cref{fig:temp}. It can be seen that the indoor temperature lies within the range of 20-22°C during the month and fluctuates with different day and night settings, regardless of whether a gas boiler or heat pump was used. However, the indoor temperature occasionally exceeds this range, which is due to the influence of factors such as delay elements (e.g., the liquid model) and the step size setting (i.e., simulation accuracy). It is clear that this model of the heating system is not ideal, and the delay properties and small amount of fluctuation present in the results make the model more similar to the real world compared to an ideal model.

It can also be seen from~\Cref{subfig:temp_weather} that the indoor temperature will be slightly lower than the indoor temperature in normal weather when a cold wave comes, which is also in line with common sense. An interesting point is that from~\Cref{subfig:temp_GBHP} it can be distinguished that the speed and sensitivity of temperature control is higher in the building with gas boiler and buffer tank than in the building with heat pump. This is due to the fact that whenever the house needs to be heated, the buffer tank can release its stored heat immediately without additional heating time, which allows the thermal needs of the house to be met faster and smoother. This shows the important role of the heat stored in the buffer tank in the regulation of indoor temperature.

\subsection{Natural Gas and Electrical Consumption}
\label{subsec:consumption}
\begin{figure}[ht]
     \centering
     \subcaptionbox{Natural gas consumption in building B1O1W1 (red), B1O1W2 (blue) and B4O1W2 (orange) in $m^3$.\label{subfig:ng_weather and insulation}}{
         \includegraphics[width=0.47\textwidth]{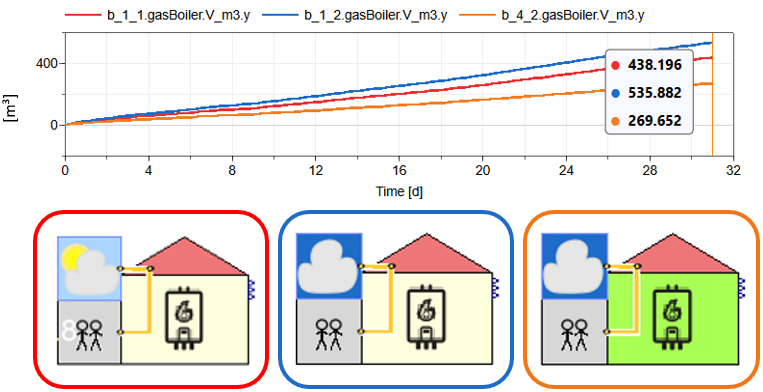}
     }
     \hspace{5pt}
     \subcaptionbox{Natural gas consumption in building B1O2W1 (pink) and B1O3W1 (cyan) in $m^3$.\label{subfig:ng_occ}}{
         \includegraphics[width=0.47\textwidth]{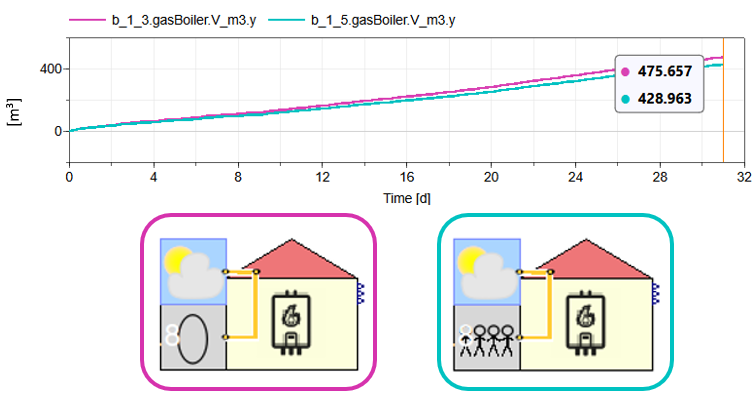}
     }
     \caption{Impact of outdoor temperature, building envelope insulation, and occupancy on natural gas consumption in January.}
     \label{fig:NG}
\end{figure}

The amount of natural gas consumed can be affected by various factors in buildings equipped with gas boilers that are fueled with natural gas. In the present paper, the authors use the higher heating value of natural gas, which is 55.5$MJ/kg$, for the calculations, with $CO_2$ emissions at combustion of 2.23$kg/kg$ and a density of 0.84$kg/m^3$. The boiler used has the same performance curve as Lochinvar's FTX500 model (\textit{lochinvar.com}), and the capacity and heating power are scaled equally based on the average heat demand of the building \citep{KB2020}. For different scenarios, the authors perform several sets of comparisons of the results for natural gas consumption, as shown in~\Cref{fig:NG}.

The comparison of buildings \texttt{\small B1O1W1} and \texttt{\small B1O1W2}  from~\Cref{subfig:ng_weather and insulation} shows that, when the outdoor temperature is low, the consumption of natural gas increases from 438.2$m^3$ to 535.9$m^3$. However, when the insulation of the building is reinforced, the consumption of natural gas decreases by approximately half compared with the initial insulated state, that is, it decreases to 269.7$m^3$. This demonstrates that the envelope modernization of a building can significantly improve its energy efficiency while maintaining resident comfort.~\Cref{subfig:ng_occ} shows that when residents are at home and have appliances operating, the generated heat can be used as a supplement to the heating system, thus consuming less natural gas. This conclusion further validates the comparability of the model with the real physical world. However, in reality, if no one is home, the heating system generally shuts down or simply maintains the room temperature at a lower level. Owing to space constraints, this situation is not considered in the present paper.

\begin{figure}[htb]
{
\centering
\includegraphics[width=0.65\textwidth]{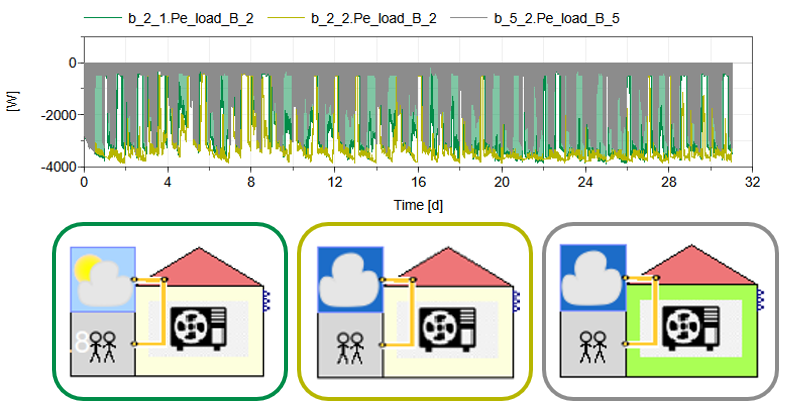}
\caption{Electrical power consumption (negative value) in three buildings  B2O1W1 (green), B2O1W2 (yellow), and B5O1W2 (grey) equipped with heat pumps under different conditions of outdoor temperature and building envelope insulation.}\label{fig:Pe weather insulation}
}
\end{figure}

Unlike boilers, heat pumps consume electricity instead of natural gas when providing heat. The authors conduct a comparative analysis of three buildings \texttt{\small B2O1W1}, \texttt{\small B2O1W2}, and \texttt{\small B5O1W2} equipped with heat pumps to demonstrate the impact of each variable on the electricity consumption of the house, as shown in~\Cref{fig:Pe weather insulation}. The apparent conclusion is similar to that in~\Cref{subfig:ng_weather and insulation}, where the unmodernized building consumes the highest amount of electricity during cold waves, and the modernized building is the most environmentally friendly.

\subsection{Overall Comparison of Results}
In this section, the authors list, compare, analyze, and summarize the simulation results for all 36 scenarios, considering \ac{PV} power generation, battery storage, natural gas costs, electricity purchase and sale costs, and $CO_2$ emissions. The parameters of natural gas used for heating the house are mentioned in~\Cref{subsec:consumption}, from which the $CO_2$ emissions of natural gas can be deduced as $Emi_{gas} = V_{gas} \cdot d_{gas} \cdot 0.00223$ in $t$. The $CO_2$ emission factor of the German power mix in 2023 is approximately 380$g/kWh$ \citep{StatistaCO22024}, which is used to calculate the equivalent $CO_2$ emissions produced by the electricity used in the house. In terms of energy price, the heating cost is divided into fuel and electricity costs. The local municipal utilities announced that the basic electricity price is 39.86$ct/kWh$ from April 2024 (\textit{stadtwerke-bruchsal.de}), and the spot market price of natural gas in Germany in April averaged 29.07\euro$/MWh$ \citep{StatistaGas2024}. Based on these assumptions, the energy consumption, $CO_2$ emissions, and heating costs of all the simulation scenarios are summarized in~\Cref{tab:results}.
\begin{table}[htb]
\caption{Natural gas consumption, electricity consumption, generation and demand from the grid, $CO_2$ emissions and heating costs for each scenario in January.}\label{tab:results}
\centering
\begin{tabular}{ccccc|cc}
\hline
Scenario Code & $V_{gas}$ [$m^3$] & $E_{load}$ [$kWh$] & $E_{gen}$ [$kWh$] & $E_{feed}$ [$kWh$] & $CO_{2,emi}$ [$t$] & Costs [\euro] \\ \hline
B1O1W1 & 438.195 & -296.523 & - & -296.523 & \cellcolor{red!40}0.934 & \cellcolor{green!10}283.155 \\
B1O1W2 & 535.885 & -296.523 & - & -296.523 & \cellcolor{red!40}1.116 & \cellcolor{green!10}319.932 \\
B1O2W1 & 475.661 & -54.351 & - & -54.351 & \cellcolor{red!40}0.912 & \cellcolor{green!10}200.73 \\
B1O2W2 & 571.857 & -54.351 & - & -54.351 & \cellcolor{red!40}1.092 & \cellcolor{green!10}236.944 \\
B1O3W1 & 428.966 & -361.164 & - & -361.164 & \cellcolor{red!40}0.941 & \cellcolor{green!10}305.447 \\
B1O3W2 & 526.779 & -361.164 & - & -361.164 & \cellcolor{red!40}1.124 & \cellcolor{green!10}342.269 \\
B2O1W1 & - & -1856.21 & - & -1856.21 & \cellcolor{red!10}0.705 & \cellcolor{red!40}739.885 \\
B2O1W2 & - & -2258.76 & -& -2258.76 & \cellcolor{red!10}0.858 & \cellcolor{red!40}900.342 \\
B2O2W1 & - & -1768.41 & - & -1768.41 & \cellcolor{red!10}0.672 & \cellcolor{red!40}704.888 \\
B2O2W2 & - & -2127.91 & - & -2127.91 & \cellcolor{red!10}0.809 & \cellcolor{red!40}848.185 \\
B2O3W1 & - & -1883.27 & - & -1883.27 & \cellcolor{red!10}0.716 & \cellcolor{red!40}750.671 \\
B2O3W2 & - & -2287.95 & - & -2287.95 & \cellcolor{red!10}0.869 & \cellcolor{red!40}911.977 \\
B3O1W1 & - & -1856.21 & 404.9 & -1464.91 & \cellcolor{green!10}0.557 & \cellcolor{red!10}583.913 \\
B3O1W2 & - & -2258.76 & 404.9 & -1858.47 & \cellcolor{red!10}0.706 & \cellcolor{red!40}740.786 \\
B3O2W1 & - & -1768.41 & 404.9 & -1378.16 & \cellcolor{green!10}0.524 & \cellcolor{red!10}549.335 \\
B3O2W2 & - & -2127.91 & 404.9 & -1728.83 & \cellcolor{red!10}0.657 & \cellcolor{red!40}689.112 \\
B3O3W1 & - & -1883.27 & 404.9 & -1492.36 & \cellcolor{green!10}0.567 & \cellcolor{red!10}594.855 \\
B3O3W2 & - & -2287.95 & 404.9 & -1887.84 & \cellcolor{red!10}0.717 & \cellcolor{red!40}752.493 \\
B4O1W1 & 212.111 & -296.523 & - & -296.523 & \cellcolor{green!10}0.51 & \cellcolor{green!40}198.045 \\
B4O1W2 & 269.661 & -296.523 & - & -296.523 & \cellcolor{red!10}0.618 & \cellcolor{green!10}219.71 \\
B4O2W1 & 254.625 & -54.351 & - & -54.351 & \cellcolor{green!10}0.498 & \cellcolor{green!40}117.52 \\
B4O2W2 & 311.144 & -54.351 & - & -54.351 & \cellcolor{red!10}0.603 & \cellcolor{green!40}138.796 \\
B4O3W1 & 203.938 & -361.164 & - & -361.164 & \cellcolor{green!10}0.519 & \cellcolor{green!10}220.734 \\
B4O3W2 & 261.63 & -361.164 & - & -361.164 & \cellcolor{red!10}0.627 & \cellcolor{green!10}242.452 \\
B5O1W1 & - & -973.272 & - & -973.272 & \cellcolor{green!10}0.37 & \cellcolor{green!10}387.946 \\
B5O1W2 & - & -1186.33 & - & -1186.33 & \cellcolor{green!10}0.451 & \cellcolor{red!10}472.871 \\
B5O2W1 & - & -883.185 & - & -883.185 & \cellcolor{green!10}0.336 & \cellcolor{green!10}352.038 \\
B5O2W2 & - & -1098.12 & - & -1098.12 & \cellcolor{green!10}0.417 & \cellcolor{red!10}437.711 \\
B5O3W1 & - & -1003.04 & - & -1003.04 & \cellcolor{green!10}0.381 & \cellcolor{green!10}399.812 \\
B5O3W2 & - & -1218.42 & - & -1218.42 & \cellcolor{green!10}0.463 & \cellcolor{red!10}485.662 \\
B6O1W1 & - & -973.272 & 404.9 & -605.212 & \cellcolor{green!40}0.23 & \cellcolor{green!10}241.238 \\
B6O1W2 & - & -1186.33 & 404.9 & -810.742 & \cellcolor{green!10}0.308 & \cellcolor{green!10}323.162 \\
B6O2W1 & - & -883.185 & 404.9 & -515.471 & \cellcolor{green!40}0.196 & \cellcolor{green!10}205.467 \\
B6O2W2 & - & -1098.12 & 404.9 & -723.219 & \cellcolor{green!40}0.275 & \cellcolor{green!10}288.275 \\
B6O3W1 & - & -1003.04 & 404.9 & -634.085 & \cellcolor{green!40}0.241 & \cellcolor{green!10}252.746 \\
B6O3W2 & - & -1218.42 & 404.9 & -843.771 & \cellcolor{green!10}0.321 & \cellcolor{green!10}336.327 \\
\hline
\end{tabular}
\end{table}

To visualize the results, the data in the $CO_2$ emissions and heating costs columns are marked with different colors. Lower emissions and costs are labeled green, followed by light green, and vice versa in red and light red. It can be seen that poorly insulated buildings heated by gas boilers (B1) have the highest $CO_2$ emissions, but are relatively inexpensive to heat. However, installing heat pumps in such buildings (B2 and B3), although it reduces emissions, results in high heating costs, and even if the cost of electricity can be partially compensated by installing a \ac{PV} battery system, the high cost of such a system is not a cost-effective solution.

In contrast, the use of heat pumps and \ac{PV} systems in renovated buildings (B6) not only significantly reduces greenhouse gas emissions but also has significant economic advantages. In addition, although the building variant with the lowest heating costs is B4, it does not meet the trend of sustainability due to its high carbon footprint. In summary, if the acquisition costs of heat pumps are not considered, heat pumps are an ideal heating device in terms of sustainability and economy in well-insulated buildings, whereas in older buildings, the use of gas boilers for heating is an economical decision.

\subsection{Discussion}
From the power generation data $E_{gen}$ in~\Cref{tab:results}, it can be seen that the \ac{PV} power generation is lower than the load in this month, but the existence of the \ac{PV} system still reduces the carbon emissions and heating costs to a certain extent. In fact, \ac{PV} generation is related to factors such as the area, angle of the photovoltaic panel, outdoor temperature, and sunlight intensity etc. Therefore, optimization control of the \ac{PV} system can effectively improve the sustainability and economy of energy use.

In buildings where gas boilers are used for heating, fuels other than natural gas are not analyzed for comparison in the present paper. Natural gas has a significant price advantage over hydrogen. However, with technological progress, a reduction in the cost of hydrogen production, transportation, and storage can make hydrogen-fueled heating with boilers a sustainable and economical option with great potential for the future.

\section{Conclusion and Outlook}
\label{sec:conclusion}
In this paper, a configurable modular multi-modal generic white-box building model is presented, which contains not only a building envelope model and electrical and thermodynamic elements, but also interfaces to occupancy models and weather models. With the cooperation of the modules, the simulation and analysis of the thermodynamic behavior of the building, as well as sustainability and economic analysis, are realized. The results show that heat pump heating has a significant advantage over natural gas heating in well-insulated buildings. However, for poorly insulated buildings, heat pumps can incur a high electricity cost. This suggests that not only is the popularization of heat pumps important, but also the upgrading of the insulation of houses is necessary to improve the energy efficiency of society and to promote the concept of sustainable development.

In addition, all modules of the introduced model can be turned on and off, and parameters can be modified by using parameter tables, which greatly simplifies the automation of generating building models with different configurations, and provides the underlying foundation for large-scale co-simulation of districts and even cities in the future. This modular framework demonstrates its potential for comprehensive scenario analysis and offers valuable insights into the sustainability of building energy systems under different conditions. More importantly, both the electrical and thermodynamic modules are configured with interfaces for coupling with the power, heat, and even natural gas grid, such as the power and gas demand of the house, the temperature and flow demand of the hot water, etc. Thus, in a large-scale simulation, this generic building model can be used to form an integrated co-simulation model with multiple energy grid models, which contributes to the digital twinning of the energy system.

\bibliographystyle{unsrtnat}
\bibliography{main}  %%% Uncomment this line and comment out the ``thebibliography'' section below to use the external .bib file (using bibtex) .





% \end{thebibliography}

\end{document}